\documentstyle[prd,aps,floats,epsf]{revtex}

\newcommand{\be}{\begin{equation}}
\newcommand{\ee}{\end{equation}}

\begin{document}

\draft

%
%

%
\renewcommand{\topfraction}{0.99}
\renewcommand{\bottomfraction}{0.99}

\twocolumn[\hsize\textwidth\columnwidth\hsize\csname 
@twocolumnfalse\endcsname

\bibliographystyle{unsrt}

\title{Nonsingular Dilaton Cosmology in the String Frame}

\author{Damien A. Easson$^{1}$ and Robert H. Brandenberger$^{1}$ }
\address{Department of Physics,  Brown University, 
Providence, RI 02912, USA.}

\maketitle
\begin{abstract}

We consider the theory obtained by adding to the usual string frame dilaton gravity action specially constructed higher derivative terms motivated by the limited curvature construction of \cite{MukhanovET1992a} and determine the spatially homogeneous and isotropic solutions to the resulting equations of motion.  All solutions of the resulting theory of gravity with these symmetries are nonsingular and all curvature invariants are bounded. For initial conditions inspired by the pre-big-bang scenario solutions exist which correspond to a spatially flat  Universe starting in a dilaton-dominated superinflationary phase with ${\dot H} > 0$ and having a smooth transition to an expanding Friedmann Universe with ${\dot H} < 0$. Hence, the graceful exit problem of pre-big-bang
cosmology is solved in a natural way.

\vspace{10pt}
BROWN-HET-1173 \\
JHEP 9909:003, 1999
\vspace{10pt}

\end{abstract}

]


\section{Introduction}  

One of the paramount problems manifest in modern cosmological theories 
is that of the initial singularity.  Gasperini and Veneziano have proposed
the pre-big-bang (PBB) model$^{\cite{GasperiniET1992b}}$ based on the 
assumption that the Universe evolved out
of a ten-dimensional Minkowski space-time which is an exact vacuum of string
theory, in hopes of resolving the singularity problem.  Recently, Brandenberger,
Easther and Maia$^{\cite{BrandenbergerET1998a}}$ managed to construct 
nonsingular cosmological solutions based
on initial conditions resembling those of the pre-big-bang scenario.  They
began with dilaton gravity and the low energy effective action of string 
theory in the Einstein frame, adding to it specially constructed higher 
derivative terms.  As the string frame is the fundamental frame with
respect to
string theory$^{\cite{Veneziano1997a}}$, it is important to reconsider the analysis of \cite{BrandenbergerET1998a}, performing 
the calculations directly
in the string frame.  The Einstein frame metric $\tilde g_{\mu\nu}$ is
related to the string frame metric $g_{\mu\nu}$ via a conformal transformation
$^{\cite{Dicke1961a}}$
\be \label{eq:conft}
\tilde g_{\mu\nu} = e^{- \varphi} g_{\mu\nu}\,,
\ee
where $\varphi$ is the dilaton.

The (PBB) model is constructed from the low energy effective action of string
theory.  In the string frame, after compactification to four dimensions, the
action is given by$^{\cite{Polchinski1998a}}$
\be \label{eq:sfaction}
S = - \frac{1}{2\kappa^2}\int{d^4\!x \,\sqrt{-g} \, e^{-\varphi}
	\left\{ R + (\nabla \varphi)^2 + \cdots \right\}}\,,
\ee
where $\kappa^2 = 8 \pi G = 8 \pi m_{pl}^{-2}$.

The PBB model gives us hope of finding a nonsingular cosmology in which
in the Einstein frame the Universe starts out in a cold dilaton-dominated contracting phase,
goes through a bounce and then emerges as an expanding FRW
Universe.\footnote{See http://www.to.infn.it/$\sim$gasperin/ for an updated
collection of papers on pre-big-bang cosmology.} In the string frame this
corresponds to$^{\cite{GasperiniET1993a}}$ a Universe starting in a superinflationary phase with ${\dot H} > 0$ and emerging as a usual FRW Universe with ${\dot H} < 0$.

As is well known, it is impossible to smoothly connect the contracting and expanding
branches of the PBB model using only the tree-level action 
$^{\cite{BrusteinET1994b,EastherET1995a,KaloperET1995a,KaloperET1995b}}$.
These branches are separated by a future and past singularity of the branches respectively.
However, one-loop effects in superstring cosmology are capable of regulating the
singularities$^{\cite{EastherET1996a}}$ and smoothly connecting a
contracting phase to an expanding phase in the presence of
spatial curvature.  Refs
\cite{AntoniadisET1993a,%
LarsenET1996a,%
Rey1996a,%
GasperiniET1996b,%
GasperiniET1996c,%
GasperiniET1997a,%
Bose1997a,%
BoseET1997a,%
KalyanaRama1997a,%
DabrowskiET1997a,%
LukasET1997b,%
BrusteinET1997a,%
Maggiore1997a,%
Gasperini1998a,%
BrusteinET1997c,%
KantiET1998a}
describe other attempts to regulate the singularities of pre-big-bang
cosmology. Most of these approaches, however, have the drawback of being
perturbative in nature, and that their consequences are only felt in the
region of the theory which can only be described by the full non-perturbative physics. We wish to discuss a mechanism to solve the graceful exit problem which involves non-perturbative physics. In the absence of a non-perturbative definition of string theory we will discuss a toy model which only contains
gravity and the dilaton.

It is natural to assume that any effective theory 
of gravity obtained from string
theory, quantum gravity, or by integrating out matter fields, will
contain higher derivative terms in the action.  The question we wish to
address is if by adding specially chosen
higher derivative terms to the Ricci scalar of the Einstein action we
are able to cancel the divergences of classical gravity present at extremely 
large curvatures.

Hence, our general approach will be to construct a higher order theory
of gravity admitting cosmological solutions which are everywhere nonsingular
and compatible with the PBB scenario.

One such construction is carried out in
Refs \cite{MukhanovET1992a,BrandenbergerET1993a}. The
resulting action includes a particular combination of quadratic
invariants of the Riemann tensor, similar to a Gauss-Bonnett term
added to the usual Einstein-Hilbert action for gravity.  
The inclusion of this invariant causes all solutions of the equations
of motion to approach de Sitter space-time at high curvature, which is
naturally nonsingular. Although in general most higher-derivative theories of 
gravity have much worse singularity properties than Einstein gravity,
this particular construction stems from the ``limiting curvature hypothesis"
(which we will discuss more below) and is entirely nonsingular.

The most elementary way to formulate a nonsingular gravitational
theory is to add an
invariant $I_2$ to the action, with the property that $I_2 = 0$ is
true if and only if the spacetime is de Sitter. By coupling
$I_2$ into the gravitational action via a Lagrange multiplier field
$\psi$ with a potential chosen to ensure that $I_2 \rightarrow 0$ at
large curvatures we can force all solutions to approach de Sitter
space in this regime.  For
homogeneous and isotropic spacetimes, a simple choice for $I_2$ which
satisfies the aforementioned criteria$^{\cite{BrandenbergerET1998a}}$ is 
\be
I_2 = \sqrt{4 R_{\mu \nu} R^{\mu \nu} - R^2} \, .
\label{I2}
\ee
Recall that the physically measurable curvature invariants
such as $R$ and $R_{\mu \nu} R^{\mu \nu}$ are constructed from
the {\it string frame} metric, $g_{\mu\nu}$.
The simplest desired action including $I_2$ is
\be \label{action}
S(g_{\mu \nu}, \psi) = \int d^4\!x \sqrt{-g} \, e^{-\varphi}(R + \psi I_2 
- V(\psi))
\ee
where $V(\psi)$ is a function chosen such that the action has
the correct Einsteinian low curvature
limit for $\psi \rightarrow 0$, whereas for $|\psi| \rightarrow \infty$ 
the constraint equation forces $I_2 \rightarrow 0$. After integrating out
$\psi$, we obtain a higher derivative gravity theory which is nonpolynomial in
the curvature.

Note that introducing the Lagrange multiplier field $\psi$ is a convenient way of writing a non-polynomial higher derivative gravity action in a polynomial form. $\psi$ has no independent physical meaning. By itself, it neither represents a new matter field, nor does it stand for a particular symmetry. The situation is analogous to the construction \cite{MukhanovET1992a} of the action $S_R \, = \, \int dt \sqrt{1 - {\dot x}^2}$ for point particle motion in special relativity from the corresponding action $S_{NR} \, = \, \int dt {1 \over 2} {\dot x}^2$ for point particle motion in Newtonian mechanics. If the Lagrange multiplier construction analogous to the one employed in this paper is used to construct the new theory, with a general potential satisfying the asymptotic conditions required, then the resulting theory will have bounded velocity but will not be the theory for point particle motion in the Special Relativity, 
i.e. in the theory with the new symmetry. $S_R$ can only  
be obtained by means of a very special choice of the coefficients of the
potential for $\psi$. In analogy, we cannot expect to be able to extract any 
new stringy symmetry from the ad-hoc construction of this work.
 
In this paper, we will show that the addition of the same higher
derivative terms as used above in (\ref{action}) to the action of PBB cosmology results in the 
elimination of singularities and connects the (in the Einstein frame) dilaton driven contracting phase with the expanding phase via a bounce.  

{F}rom the point of view of string theory our construction appears, unfortunately, rather artificial in that we are picking out a particular narrow class of (nonpolynomial) higher derivative gravity actions from the infinite dimensional space of such actions. Our work should be viewed as an existence
proof (in the string frame) of a higher derivative dilaton gravity theory with nonsingular cosmological solutions rather than as a realistic string-motivated model. 

In particular, the invariant $I_2$ which plays a crucial role in our construction is not one of the curvature invariants which appears in the next to leading order in derivatives of the low energy effective action of string theory.  The role of the invariant $I_2$ and of the associated Lagrange multiplier field is to implement the limiting curvature hypothesis. There are indications that this hypothesis is realized in nonperturbative string theory (see e.g. \cite{BV}), as a consequence of t-duality. This duality is broken when considering the low energy effective action of string theory, and it is therefore not to be expected that the curvature invariants which arise in this low energy effective action can be used in place of $I_2$. The invariant $I_2$ plays the role of non-perturbative information about the theory. Obviously, since the non-perturbative definition of string theory is still unknown, we cannot hope that $I_2$ yields more than a toy model for what string theory will
eventually tell us. Since the limiting curvature hypothesis plays an important
role in both our construction and in string theory, we can hope that our model
may eventually turn out to contain some of the relevant physics.
 
We should also warn the reader that our work has nothing to say concerning the initial condition problem$^{\cite{KLB}}$ of the pre-big-bang scenario.

\section{Action and Equations of Motion}

We begin with the string frame action for dilaton gravity,
adding to it the higher derivative
term given by $I_2$, in analogy to what was done in the absence of the
dilaton in Refs \cite{MukhanovET1992a,BrandenbergerET1993a} and in the
Einstein frame in Ref \cite{BrandenbergerET1998a}:
\be  \label{E4}
S = \frac{-1}{2\kappa^2} \int{ d^4\!x \,\sqrt{-g} \, e^{-\varphi} \left\{
R + (\nabla \varphi)^2 + c \psi e^{\gamma\varphi} I_2 - V(\psi) 
\right\}}.
\ee
For the moment, we allow a general coupling between $I_2$ and the
dilaton.  Minimal coupling corresponds to setting the constant
$\gamma$ equal to zero. The constant $c$ rescales the Lagrange
multiplier field $\psi$, and will be chosen to simplify the equations
of motion.

Consider a homogeneous and isotropic metric of FRW type
\be
ds^2 = n^2(t)dt^2 - a(t)^2 \bigl( {1 \over {1 - kr^2}}dr^2 + r^2 d\Omega^2
\bigr) \, ,
\ee
where $d\Omega^2$ is the metric on $S^2$ and $n(t)$ is an arbitrary lapse function
which we will set to unity in the EOM below. The equations of motion
resulting from varying (\ref{E4}) with respect to $\varphi$, $\psi$ and $n$
become

$$\ddot{\varphi} + 3 H \dot{\varphi} - \frac{1}{2} \dot\varphi^2 + 
	\frac{(1-\gamma)}{2} c \psi e^{\gamma\varphi}\sqrt{12} 
\left(\frac{k}{a^2} - \dot{H} \right) = 0 $$
$$\dot{H} =  \frac{k}{a^2} - \frac{e^{-\gamma\varphi}}{c \sqrt{12}} 
\frac{\partial V}{\partial \psi} $$

\begin{eqnarray}
&6 \frac{k}{a^2} + 6H^2 + \dot{\varphi}^2 - 6H\dot\varphi - V(\psi) 
	= &\nonumber\\
&c e^{\gamma\varphi}\sqrt{12} \left(3 H^2 \psi -\frac{k}{a^2}\psi +
H(\dot{\psi} + (\gamma-1) \dot{\varphi}\psi)\right), & \label{E6}
\end{eqnarray}
respectively, where dots denote differentiation with respect to time, $t$.

For a spatially flat, bouncing
Universe, we set the curvature constant $k = 0$. We will simplify things
further by considering minimal coupling of $\varphi$ to $I_2$, i.e.
$\gamma = 0$. To eliminate useless constant coefficients in the
equations of motion, it is convenient to choose $c = 1/\sqrt{12} $. The
resulting equations of motion become
\begin{eqnarray} \label{EOM}
\dot{\psi} \, &=& \, - 3 H \psi \, + 6 H \, + (\chi - 6)\psi \,+ \, {1 \over H}
\bigl( \chi^2 \, - \, V(\psi) \bigr), \nonumber \\
\dot{H} \, &=& \, \ -V^{\prime}(\psi) , \nonumber \\
\dot{\chi} \, &=& \, - 3 H \chi + \frac{1}{2}(\chi^2 + \psi\dot H),  \label{E7}
\end{eqnarray}
with $\chi = \dot{\varphi}$ and the prime ($\prime$) denoting differentiation
with respect to $\psi$.
	
Now let us focus on the construction of the potential
$V(\psi)$.  Once again, our arguments are the same as in 
Ref \cite{BrandenbergerET1998a}.  When the curvature is small, the terms 
in the action
(\ref{E4}) with Lagrange multiplier field $\psi$ dependence must be negligible 
compared to the usual terms of dilaton gravity. This is made manifest by 
the condition
\be  
V(\psi) \, \sim \, \psi^2 \,\,\,\,\,\, |\psi| \rightarrow 0 \label{E8}
\ee
since the region of small $|\psi|$ will correspond to the low curvature
regime.$^{\cite{BrandenbergerET1993a}}$ In order to implement the
limiting curvature hypothesis, the invariant $I_2$ must tend to zero,
and hence the metric $g_{\mu \nu}$ will approach a de Sitter metric at
large curvatures, i.e. for $|\psi| \rightarrow \infty$.  From the
variational equation with respect to $\psi$, we find the requirement
\be
V(\psi) \, \rightarrow \, {\rm const} \,\,\,\,\,\, |\psi| \rightarrow
\infty \, . \label{E9}
\ee

As in \cite{BrandenbergerET1998a} we add a third condition in order to ensure that there is a bouncing solution for $k=0$. The equations must allow a configuration with $H = 0$ and $\psi \neq 0$. From the equation of motion for $\psi$ in (\ref{E7}) (and considering the region of small $\chi$) it follows that $V(\psi)$ must become negative, assuming that it is positive for small $|\psi|$. Let $\psi_b$ denote the value of $\psi$ at the bounce.  This will occur when
\be V(\psi_b) \, = \, \chi^2  \, , \label{E10}
\ee
as can be seen from the equation of motion for $\psi$ setting
$H = 0$.

A simple potential which satisfies the conditions (\ref{E8}), (\ref{E9})
and goes negative beyond some critical value of $\psi$ is
\be
V(\psi) = \frac{\psi^2 - \frac{1}{16}\psi^4}{1+\frac{1}{32} \psi^4}.
\ee
Note that the specific values of the coefficients of the terms in $V(\psi)$ are not important as long as the three criteria discussed above are satisfied. 

\section{Phase Diagram of Solutions}

The condition (\ref{E10}) on the potential $V(\psi)$
ensures that our model is nonsingular and geodesically complete for large values of $|\psi|$ but says nothing about geodesic completeness of 
solutions which always remain within the small $|\psi|$ regime.  Hence
it is necessary to study the small $|\psi|$ region in order to prove
that our model is everywhere nonsingular.

In this section we discuss the projection of the three dimensional phase space
$(\psi, H, \chi)$ of our model into the two dimensional 
phase plane $(\psi, H)$ for small values of $\chi$.  By projecting out the $\chi$ axis we lose no relevant
information about the system since the curvature invariants are not affected by $\chi$.  Both analytical and numerical methods are used to
study the trajectories of solutions of (\ref{E7}) in the phase plane
and thus explicitly demonstrate the absence of singularities.  Furthermore, and
unlike in the model of Refs. \cite{MukhanovET1992a,BrandenbergerET1993a},
we shall show that our theory admits spatially flat bouncing
solutions as in the Einstein frame model of \cite{BrandenbergerET1998a}.

We need to demonstrate that all solutions tend to a known nonsingular and geodesically complete space-time at large positive and negative times. With the symmetries of our problem, these asymptotic space-times are either Minkowski or de Sitter. We need to show that all phase space trajectories either tend to the origin of phase space or else to $|\psi| \rightarrow \infty$ for finite values of $H$. We will first analyze the phase space trajectories for small values of $|\chi|$. In Section 4 we then discuss the case of large $|\chi|$.

There are many interesting points and curves on the phase plane $(\psi, H)$ which demand special attention.
First, the point
$(\psi, H, \chi) = (0, 0, 0)$ corresponds to Minkowski space-time.
The potential $V(\psi)$ vanishes at this point, but it also vanishes at
the values 
\be
\psi_b = \pm 4 \, .
\ee
{}From the first equation in (\ref{EOM}) it follows that the phase plane points $(\psi_b, 0,0)$ correspond
to bouncing points of cosmological trajectories. The general equation for a possible bounce is 
\be
V(\psi) \, = \, \chi^2 \, .
\ee
To prove that this is in fact a bounce, we need to show that contracting solutions are attracted to it. This will be done below.

$\dot H$ vanishes when the derivative of $V$ does, i.e. at the values
\be\label{eq:psid}
\psi_d = \pm 2 \,.
\ee
Trajectories which cross the phase space planes $(\psi_d, H, \chi)$ have $\dot H = 0$ at the points where they cross.

To show that the curves through $(\psi, H, \chi) = (4 - f(\chi), 0, \chi)$, with $f(\chi)$ chosen such that $(\chi^2 - V) = 0$, represent
bounce solutions, we expand the $\psi$ equation of motion near
$H = 0$, which gives
\be \label{eq:bnce}
H ({\dot \psi} + 6 \psi - \chi \psi) \, = \, \chi^2 - V \,.
\ee
First note that contracting solutions with $|H| \ll 1$ and
$2< \psi < 4 - f(\chi)$ have $\dot \psi > 0$ and approach the bounce $(4 - f(\chi), 0, \chi)$
in finite time since $\dot H$ is positive and does not go to zero. If the trajectories reach $\psi > 4 - f(\chi)$ before hitting the $\psi$ axis, then ${\dot \psi}$ changes sign. This shows that the bounce is an attractor on the contracting branch. By taking the time derivative of the $\psi$ equation in (\ref{EOM}) and using the other two equations to eliminate $\dot \chi$ and $\dot H$ and after expanding near the bounce it can be verified that on the expanding side of the branch $\ddot \psi < 0$ and that therefore the trajectories will eventually turn around (i.e. $\dot \psi < 0$) (see Figure 1).

\begin{figure}[tbp]
\begin{center}
\begin{tabular}{c}
\epsfxsize=8cm 
\epsfbox{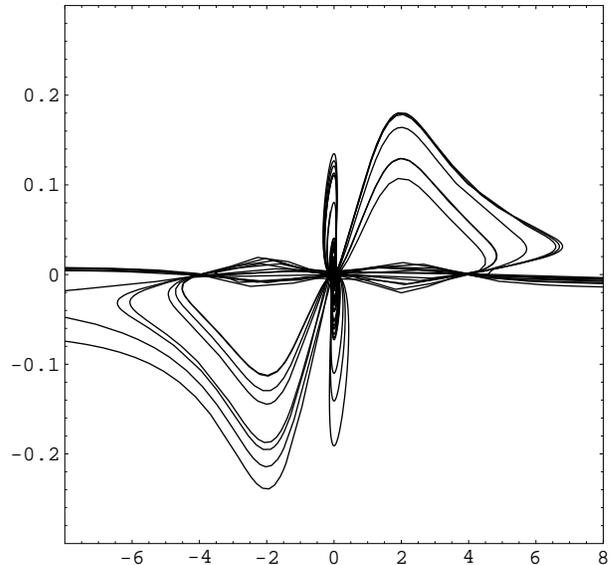}
\end{tabular}
\end{center}
\caption[fig1]{Projection of a set of trajectories onto the $\psi$ - $H$ plane of phase space for small values of $|H|$ and $\chi$. The Hubble parameter, $H$, is plotted on the vertical axis, while $\psi$ is plotted on the horizontal. Shown are a set of solutions which oscillate about Minkowski space-time $H = 0$, $\psi = 0$ as well as solutions through the bounce points $H = 0$ and $\psi \simeq \pm 4$.}
\end{figure}
  
By expanding the equations about the origin we find that
\be\label{eq:critp}
\frac{d\psi}{dH} \simeq \frac{1}{2H}(\psi - \frac{6H^2 + \chi^2}{\psi} + 6H)
\ee
which defines the critical surface
\be \label{csurf}
\psi^2 + 6 \psi H - 6H^2 - \chi^2 = 0
\ee
Note that lines on the critical surface are not phase space trajectories of
the equations of motion.  Such trajectories would point in the vertical
direction at the critical surface since $d\psi/dH =0$. Note also that as $\chi^2$ increases, the critical surface approaches the $\psi$ axis.

{}For the contracting solutions we find that the equation of motion
for $\dot H$ becomes
\be \label{eq:eoh}
\dot H \simeq -2 \psi \, .
\ee
Hence contracting solutions with $\psi > 0$ which are above the critical surface have $\dot H<0$ and from Eq. (\ref{eq:critp}) $\dot \psi > 0$.  These
trajectories are pulled toward the $\psi = 2$ curve where $\dot H$
changes sign.  Trajectories which start out in this region, do
not cross the critical surface and reach the $\psi = 2$ surface at a value $|H| \ll 1$ are good candidates for spatially flat bouncing Universes (see Figure 1).  The analysis of trajectories through $(\psi, H, \chi) = (- 4 + f(\chi), 0, \chi)$ is similar. Trajectories near the origin below the critical surface correspond to solution which oscillate about Minkowski space-time (see Figure 1).

There is a separatrix surface in the lower right quadrangle of the phase plane which separates trajectories which reach $\psi = 2$ from those which do not and which eventually cross the critical line. The solutions for trajectories above this separatrix line close to the origin can be obtained by taking the time derivative of the second equation in (\ref{EOM}) and inserting the first equation and by considering terms which dominate near the origin. For small values of $\chi^2$ we obtain
\be
H {\ddot H} \, = \, - 6 {\dot H} H + {1 \over 2} {\dot H}^2 \, ,
\ee
which is the same equation as the corresponding one in
\cite{BrandenbergerET1998a} except for the first term on the right hand side.
However, by inserting as an ansatz the solutions found in \cite{BrandenbergerET1998a}
\be
H(t) \, = \, - c t^2 
\ee
where $c$ is a constant, it is not hard to show that for small values of $t$, the extra term in our equation gives negligible corrections to these trajectories. For sufficiently small values of $c$, the trajectories lead
to bouncing solutions, whereas for somewhat larger values of $c$, the trajectories cross the $\psi = 2$ plane at $|H| > 1$. 

{}For the latter solutions, the $\psi$ equation of motion for $\psi \gg 2$ becomes
\be
{\dot \psi} \, = \, - (3 H + 6) \psi + {1 \over H} \chi^2 \, .
\ee
{}For large values of $|H|$ and not too large values of $\chi^2$, ${\dot \psi} > 0$ and $H \rightarrow 0$ and the trajectory tends to de Sitter space. However, as will be discussed in Section 4, $\chi$ is rapidly growing, and eventually (unless the dilaton is stabilized), the $\chi^2$ term in the above equation takes over and leads to ${\dot \psi} < 0$, in which case these solutions also tend to de Sitter space, but at large negative values of $\psi$. The bottom line is that there are no singularities in the phase space region $\psi > 2$ and $H < 0$.  

{}From Figure 2 we see a potentially troublesome feature
of the phase space near $\psi = 2$. There are critical lines with ${\dot \psi} = 0$ which converge to $\psi = 2$ and $H = \infty$ from the left, and to $\psi = 2$ and $H = - \infty$ from the right. Along these critical lines, the trajectories appear to head
off to infinity (for negative $H$) or emerge from infinity (for positive $H$), suggesting a singularity is present in the model.  
Upon a more detailed analysis of the EOM it becomes clear that the
solutions in this region are repelled by the critical line.  To see
this
begin with the critical line obtained by setting $\dot \psi = 0$.  Considering
the lower right quadrant of the phase plane, to the left of the critical
line, we have
\be
\psi = \frac{6 H^2 + \chi^2 - V}{H(3H + 6 - \chi)} \, .
\ee
The critical line for small values of $\chi^2$ approaches the surface $\psi = 2$ from the right (note that in the Einstein frame model of \cite{BrandenbergerET1998a} it approached from the left, the difference being due to the extra term in the $\psi$ equation of motion in (\ref{EOM})). Differentiating the $\psi$ equation of motion with respect to time and evaluating the result along the critical line we see that
\be
\ddot \psi < 0
\ee
and  hence the  
critical line is a repeller, sending the trajectories away from the 
line and toward the asymptotic de Sitter region.  To the right of
the critical line $\dot \psi > 0$ and the trajectories are again pushed
away from $\psi = 2$.

\begin{figure}[tbp]
\begin{center}
\begin{tabular}{c}
\epsfxsize=8cm 
\epsfbox{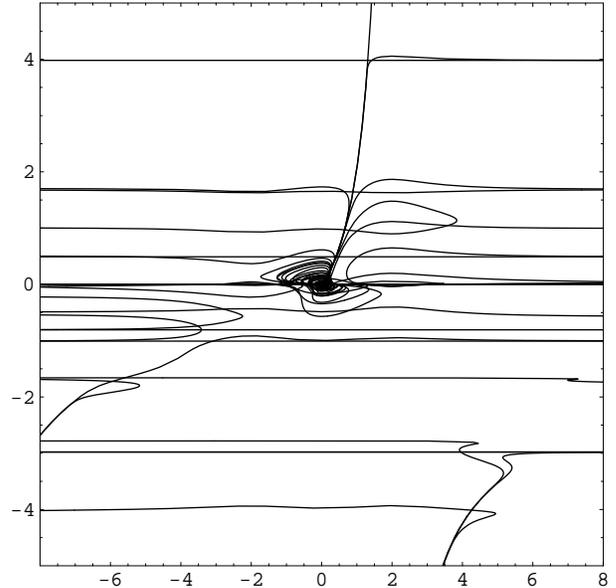}
\end{tabular}
\end{center}
\caption[fig2]{Projection of a set of trajectories onto the $\psi$ - $H$ plane of phase space for larger values of $|H|$. The axes are the same as in Fig. 1. The only regions in phase space where trajectories tend to $|H| = \infty$, i.e. to curvature singularities, are near the two critical lines which tend to $\psi = 2$. As shown in the text, these lines repel trajectories, and hence there are no singular trajectories even in these regions.}
\end{figure}

Consider the example of a frozen dilaton, $\chi \rightarrow 0$.
Hence, for  $|H| >> |\psi|$ we find 
\be
\ddot \psi \simeq (V' + 3H^2)(\psi - 2)
\ee
and $\ddot \psi < 0 $ as expected.
Solutions to the left of the critical line peel away to the 
de Sitter region while those to the right of the critical line
approach from de Sitter.

We also wish to show that $|H| \rightarrow \infty$ if and only if
$\psi \rightarrow \infty$.  For this we will consider the specific
example of the small negative $\psi$, large negative $H$ behavior.
In this region the change in $\psi$ with respect to $H$
is given by
\be\label{eq:sheri6310777}
\frac{d\psi}{dH} \simeq \frac{3H(\psi - 2)}{V'} \, .
\ee
Hence we see that the slope of the solutions increase and begin to level off
as $|H|$ increases.

To conclude this section, we have shown at this point that there are no
singularities for small values of $\chi^2$ and that all solutions can be continued to arbitrary large proper time, which demonstrates geodesic completeness. In this following section we turn to the discussion of the
dynamics for large $\chi^2$.

\section{Dilaton Evolution and Stabilization} \label{dilstab}

The dilaton is the massless scalar field with gravitational
strength couplings, found in all perturbative string theories.
It is generally believed that dynamical effects will generate
a mass for the dilaton in vacua with broken  supersymmetry  
$^{\cite{Polchinski1998a}}$. If the dilaton were to
remain massless$^{\cite{DamourET1996a}}$ it would affect the
values of gauge couplings,  and produce potentially observable
consequences.

In our model there are singularities in $\chi$.  Consider
the EOM for $\chi$ in the large $\chi^2$ limit
\be\label{eq:chisin}
\dot\chi = \frac{1}{2} \chi^2 \, .
\ee
This has solution
\be\label{eq:solu} 
\chi(t) = \frac{1}{\chi^{-1}(t_0) - \frac{1}{2}(t - t_0)} \,,
\ee
which blows up when
\be\label{eq:blow}
\chi(t_0) = \frac{2}{(t-t_0)} \,.
\ee

This singularity does not concern us however, since it can 
be avoided by adding a simple potential $U(\varphi)$ to the
dilaton equation in order to freeze the dilaton in a manner consistent
with the above argument. Where this potential comes from is the usual
problem of any theory with a dilaton, a problem to which we have no answer
either.

What happens to the projection of the phase space trajectories onto the $\psi - H$ plane in the large $\chi$ regime?  Here we have
\be\label{eq:lgchi} 
\dot \psi= \chi \psi + \frac{\chi^2}{H}
\ee 
and the $H$ equation of motion remains the same.  Immediately
we see that for $H \ne 0$, $|\dot \psi| \rightarrow \infty$ which
implies that $\psi \rightarrow \infty$ and hence $H \rightarrow$ const.
Further insight is gained by noticing that
\begin{eqnarray} \label{eq:kate}
&\dot \psi > 0 \quad \rm{for} \quad H>0& \\
&\dot \psi < 0 \quad \rm{for} \quad H<0 &.
\end{eqnarray}
We know that $V'$ vanishes for $\psi = 0$ and $\psi = \pm 2$.
Thus
\begin{eqnarray} \label{eq:chameleon}
& \psi = 0: \quad \dot \psi = \frac{\chi^2}{H}& \\
& \psi = \pm 2: \quad \dot \psi = \pm 2\chi + \frac{\chi^2}{H}& \,.
\end{eqnarray}
Therefore, solutions above $H=0$ stretch straight (more or less) 
across the $\psi-H$ plane
from left to right, and those below the $H = 0$ line stretch from 
right to left, as can be seen in Figure 3.  

In the large $\chi^2$ limit, the equation for the critical surface ${\dot \psi} = 0$ becomes
\be
\psi \, \simeq \, - {{\chi^2} \over {3 H^2}} \, .
\ee
However, since for any solution of the equations of motion in this regime $\chi^2$ grows much faster than $\psi$, trajectories do not reach the critical surface. Instead, ${\dot H} \rightarrow 0$. This completes the proof that there are no curvature singularities for large $\chi^2$, and that the solutions asymptotically approach either Minkowksi or de Sitter.

\begin{figure}[tbp]
\begin{center}
\begin{tabular}{c}
\epsfxsize=8cm 
\epsfbox{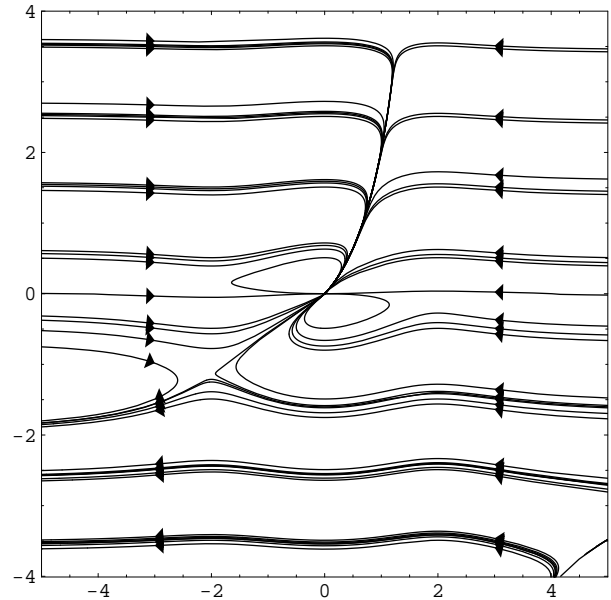}
\end{tabular}
\end{center}
\caption[fig3]{The phase portrait for solutions of the equations of
motion for fixed dilaton. The axes are the same as in Figs. 1 and 2. Due to
the extra terms in the equations of motion, the phase portrait is not symmetric under reflections about the $\psi$ axis as it was in \cite{BrandenbergerET1998a}. Clearly visible are the three critical lines. As discussed in the text, there are no singular trajectories even in the vicinity of these critical lines.}
\end{figure}

{}Finally, let us assume that the dilaton has been fixed. In this case, $\chi = 0$ and the remaining equations of motion become almost identical to the ones
studied in \cite{BrandenbergerET1998a} (for $\chi = 0$) except for the
$- 6 \psi$ term on the right hand side of the $\psi$ equation of motion in (\ref{EOM}). This term, however, is subdominant in the large $|H|$ region,
and hence the conclusions of \cite{BrandenbergerET1998a} concerning the
absence of singularities carry over to our case. The extra term does, however,
lead to asymmetries in the phase portrait under reflections about the
$\psi$ axis. There are now three critical lines where ${\dot \psi} = 0$. The first line occurs for $0 < \psi < 2$ for positive values of $H$, and with $H \rightarrow \infty$ as $\psi \rightarrow \infty$. The second line occurs for $2 < \psi < \infty$ and for negative values of $H$ with $H \rightarrow - \infty$ as $\psi \rightarrow 2$. Finally, for $\psi < 0$ there is a harmless critical line which remains at small negative values of $H$ (see Figure 3). As in Section 3 it can be shown that the two first critical lines repel the trajectories. Hence, the proof of non-singularity of the model carries over.

\section{Conclusions}

We have extended the model of nonsingular dilaton cosmology
presented in \cite{BrandenbergerET1998a} to the string frame.
Using the limiting curvature construction of \cite{MukhanovET1992a,BrandenbergerET1993a} applied to dilaton
gravity formulated in the string frame, we have obtained spatially flat 
bouncing cosmological solutions. The construction consists of adding
specially chosen higher derivative gravity terms in the form of a
curvature invariant $I_2$ to the string frame action.
The invariant $I_2$ is made up of invariants quadratic in
the Riemann curvature and has the property that $I_2 = 0$ singles out the
maximally symmetric de Sitter space-times among all homogeneous and isotropic solutions. It is coupled to dilaton gravity via
a Lagrange multiplier field $\psi$.  The $\psi$ field is nondynamical but
has a potential $V(\psi)$, which was 
chosen to allow non-singular bouncing solutions.

The three dimensional phase space of trajectories was studied
both analytically and numerically to demonstrate that all
solutions are nonsingular.  Specifically we studied a large
class of solutions which lead to bouncing cosmologies.
The dynamics of the bounce are governed by the higher derivative
gravity terms introduced by the limiting curvature construction.

The connection with pre-big-bang cosmology appears in a different form than in the Einstein frame. In the string frame picture of pre-big-bang cosmology, the Universe starts in a superexponentially expanding dilaton-dominated phase with $H > 0$ and ${\dot H} > 0$, i.e. in the upper left quadrangle of the projected phase space of Figures 2 and 3. In the absence of the higher derivate terms, the trajectories would diverge to $H = \infty$. However, as is obvious from Figures 2 and 3, the new terms we have added lead to a graceful exit from this phase. The trajectories cross the $\psi = 0$ axis and go over to trajectories which are like the usual expanding FRW solutions with $H > 0$ and ${\dot H} < 0$. This happens independent of whether the dilaton is frozen at late times or not.

There is a singularity in the dilaton equation of motion but it is assumed that
this problem can be solved by the introduction of a dilaton potential 
$U(\varphi)$.  Although we do not propose a specific form for
this potential, such a potential is physically necessary in all theories with a dilaton in order to generate a mass for the dilaton.  

The most obvious criticism of this model is that the higher order
terms in the action are artificially constructed rather than
derived from fundamental physics.  However it is important to
recall that our method is well motivated.  For one, all effective
theories of gravity, including those produced by string theory,
quantum gravity, or from quantizing matter fields in curved spacetime
must contain higher derivative terms in the action.  Furthermore, it
is natural to assume that physical invariants must be limited in
such theories to avoid singularities.  We have found that in both
the Einstein frame and the string frame, the method proposed here
ensures that all physical invariants remain finite, and produces
bouncing nonsingular cosmological solutions. The graceful exit problem
of pre-big-bang cosmology is solved naturally by our construction both in
the Einstein and string frames. The connection with
pre-big-bang cosmology will be explored further in a future publication.

\section*{Acknowledgments}

This work is supported in part by the DOE contract DE-FG0291ER40688 (Task A), and by a Department of Education award to Brown University under the GAANN program. DE would like to thank Richard Easther and Matthew Parry for many helpful discussions throughout this work, and in particular Richard Easther
for the use of his computer program to generate the phase portraits. DE also
acknowledges support during the summer of 1998 from a NASA Space Grant to Brown University.


\end{document}